# Calibrated Audio Steganalysis

Hamzeh Ghasemzadeh
Electrical Engineering Department
Islamic Azad University of Damavand
Tehran, Iran
hamzeh_g62@yahoo.com

Mohammad H. Kayvanrad
Biomedical Engineering Department
Amirkabir University of Technology
Tehran, Iran
mkayvanrad@aut.ac.ir

*Abstract*— Calibration is a common practice in image steganalysis for extracting prominent features. Based on the idea of re-embedding, a new set of calibrated features for audio steganalysis applications are proposed. These features are extracted from a model that has maximum deviation from human auditory system and had been specifically designed for audio steganalysis. Ability of the proposed system is tested extensively. Simulations demonstrate that the proposed method can accurately detect the presence of hidden messages even in very low embedding rates. Proposed method achieves an accuracy of 99.3% (StegHide@0.76% BPB) which is 9.5% higher than the previous R-MFCC based steganalysis method.

## I. INTRODUCTION

Recently different security services based on signal processing techniques have emerged [1, 2]. Steganography is among such trends and it is the art and science of concealing messages into innocuous-looking hosts [3]. Historically, it can be traced back to ancient Greece where waxed-tablets were used for covert communications. With the dawn of digital electronics and digital signal processing, this art has evolved into a fascinating science which has attracted attentions of many researchers.

Steganography in its modern definition has some essential components. On the sender side, it consists of an embedding algorithm ($A_{em}$), a host signal which is called cover ($c \in C$), and the intended message ($m \in M$). Sender applies $A_{em}$ to the cover and produces another signal which is called stego signal ($s \in S$). At the other end of the channel, the receiver applies another algorithm called extraction ($A_{ex}$) on the received stego signal and retrieves the intended hidden message. Equation (1) presents these definitions:

$$A_{em}(c,m) = s, \qquad A_{ex}(s) = m \qquad (1)$$

Steganalysis is the opposite side of steganography and it intends to detect the presence of hidden messages in other signals. Therefore, steganography and steganalysis are in constant battles, while the former proposes new embedding methods, the later tries to break it. Based on type of signals that steganography and steganalysis methods work on, they are divided into major categories of text, audio, image, video, and network traffics. Using a different taxonomy, steganalysis can be divided into targeted and universal methods. In the targeted methods, the steganalyst knows the embedding algorithm. Therefore, these methods have been designed specifically for detecting artifacts of a particular method. On the other hand, universal methods do not have any prior assumptions about the embedding method [4] and they are supposed to work properly for detection of any type of steganography. Reviewing steganalysis methods shows that they can also be divided into two major categories of calibrated and non-calibrated ones.

In the first category, features are extracted based on an estimation of the cover signal. These methods are commonly known as calibrated methods, a concept which section II covers in more details. The most common calibration method in audio steganalysis is to apply a noise removal process on the incoming signal. Then, discrepancy between incoming signal and the estimated cover can be quantified with some metrics. This idea was presented in [5] where distortions between two signals were quantified with audio quality metrics (AQMs). Liu et al. argued that AQMs have not been designed for steganalysis purposes and, therefore they are not suitable for steganalysis [6]. Instead they used Hausdorff distance to capture traces of steganography. Other researchers have used Gaussian mixture model (GMM) and generalized Gaussian distribution (GGD) for the same purpose [7]. Avcibas challenged the idea of estimating cover from the incoming signal [8]. It was shown that this self-referencing may diminish performance of steganalysis system.

The second category which can be called non-calibrated, extracts its features directly from the signal under scrutiny. First, Dittmann integrated steganalysis into intrusion detection systems [9]. Another work argued that steganography may change the correlation between neighboring samples; thus, residual signal from linear prediction code (LPC) were exploited for steganalysis [10]. Koçal et al. used chaos theory and extracted different non-linear measurements for steganalysis [11]. In [12] Mel-frequency Cepstrum coefficients (MFCC) were used for steganalysis of voice over IP (VoIP) applications. Finally, Ghasemzadeh et al. argued that extracting features based on the model of human auditory system (HAS) may discard most of the valuable information for steganalysis [13]. Considering this fact, the work proposed a new model that had maximum deviation from HAS and used it for feature extraction.

Comparing calibration in audio and image steganalysis reveals that calibration has not found its true value in audio steganalysis context. Specifically, almost all of calibrated methods in audio steganalysis are based on noise removal procedure. The goal of this work is to fill this gap to some extent. This paper uses a new set of calibrated features based on a model with inverse frequency resolution of HAS. Then, performance of the proposed system is evaluated on different data hiding algorithms.

The rest of this paper is organized as follows. Section II introduces calibration technique in more details. The proposed method is presented and discussed in section III. Section IV is devoted to simulations and experimental results. Discussion follows in section V and conclusion is made in section VI.

## II. CALIBRATION IN STEGANALYSIS

First, Let us investigate a trivial case. If steganalyst knows the cover signals, his performance will improve considerably. In this scenario, a simple comparison will lead to an accurate decision. Although this assumption is completely unrealistic and never holds in practical settings, it has motivated researchers to find accurate ways to estimate cover from the signal under scrutiny. Apparently, as the estimation gets better, the decision becomes more accurate. This has led to the idea of calibration in which the goal is to estimate the statistics of cover as accurately as possible. Calibration was first proposed in [14] and since then, many different calibration methods have been proposed for image steganalysis. They include re-compression [14], filtering [15], down-sampling [16], and etc.

This paper uses re-embedding method for audio calibration. To illustrate how re-embedding can improve performance of the system, two different situations are considered. Suppose $x$, $A_{em}$, and F denote the signal under scrutiny, the embedding method, and the feature extraction procedure, respectively. If $x$ is a stego signal ($x \in S$) then both $x$ and its re-embedded version ($\tilde{x} = A_{em}(x)$) would be stego signals. Thus, their features should be very similar. That is:

$$\{\tilde{x}, x\} \in S \Rightarrow F(\tilde{x}) - F(x) \approx 0 \qquad (2)$$

On the other hand, If $x$ is a cover signal ($x \in C$) then only its re-embedded version ($\tilde{x} = A_{em}(x)$) would be stego. Thus, their features should be very different. That is:

$$x \in C, \tilde{x} \in S \Rightarrow F(\tilde{x}) - F(x) \not\approx 0 \qquad (3)$$

Using pattern recognition terminology, equations (2) and (3) show that re-embedding reduces within class distances and increases between class distances; thus, it can improve discriminative property of features.

## III. PROPOSED METHOD

The proposed method uses higher order statistics (HOS) extracted from the difference between Reversed Mel Cepstrum Coefficients (R-MFCC) of calibrated version and its original one for the purpose of steganalysis.

### A. Reversed Mel Cepstrum

Cepstrum is an anagram of the word spectrum and it is defined as the inverse Fourier transform of the logarithm of the power spectrum of a signal.

$$cepst_{x(t)} = |F^{-1}\{log(|F\{x(t)\}|^2))\}|^2 \qquad (4)$$

Usually in speech related applications a modified version of cepstrum known as Mel cepstrum is used. The Mel scale has been designed to mimic certain characteristics of human ear; therefore, it has high frequency resolution in low frequency regions and low frequency resolution in the high frequency regions. Recently, a new scale was introduced in [13] that has the exact opposite properties. This new scale was named Reversed-Mel scale and it was specifically designed for steganalysis applications. If $F_S$ denotes sampling frequency of the signal, equation (5) shows the relation between a given frequency in hertz (f) and its R-Mel equivalence.

$$RMel = 1127 \times ln(1 + \frac{0.5 \times F_s - f}{700}) \qquad (5)$$

Figure 1 depicts both Mel and R-Mel scales versus hertz.

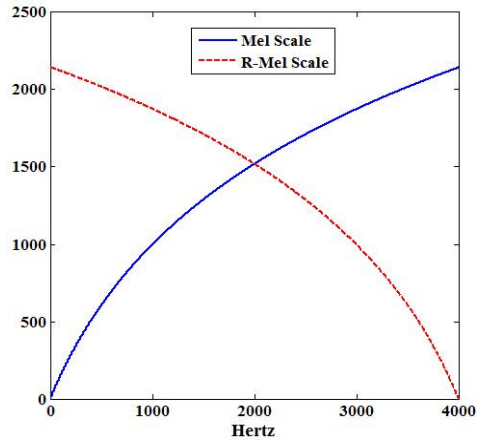

Figure 1.  Mel and R-Mel scales vs. hertz

### B. Feature Extraction

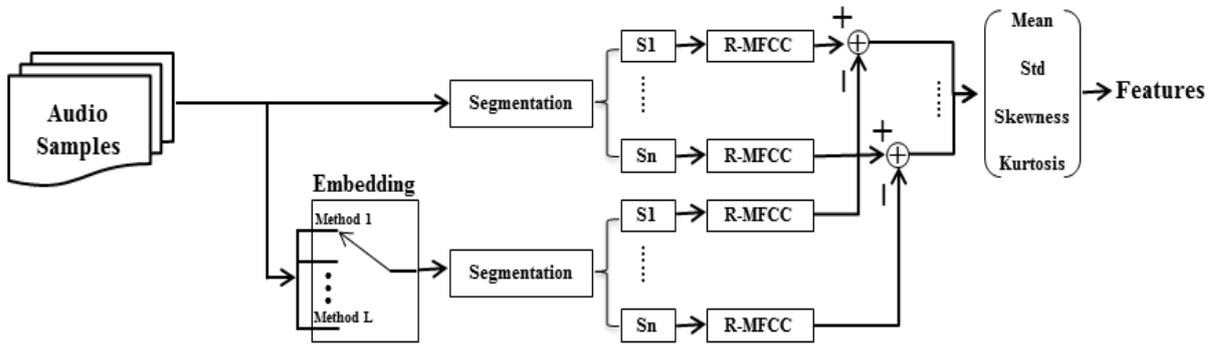

Figure 2.  Feature extraction procedure. Features are extracted as the higher order statistics of the differences between RMFCC coeficients of the incoming signal and its re-embedded version.

Liu et al. showed that taking second order derivative of audio signal can improve discriminative property of MFCCs [17]. Consequently, the features were extracted from the second order derivative of audio signals. Then data was normalized and was segmented into frames of 1024 samples with overlap value of 512 samples. Then, 29 different R-MFCCs were extracted from each segment. The same procedure was repeated for the re-embedded version of the signal. Steganalysis features were defined as the mean, standard deviation, skewness, and kurtosis of differences between R-MFCCs of original and its re-embedded version. Figure 2 shows block diagram of the proposed method. It is noteworthy that the "Embedding" block of Figure 2 contains a set of data hiding algorithms. Initially, this set was selected such that it covered all of data hiding algorithms investigated in this paper. Accordingly, in each tests the embedding algorithm was adjusted and selected according to the method under investigation.

### C. Classification:

The process of distinguishing between stego and cover signals needs a classifier to define the suitable decision boundaries. This paper employs support vector machine (SVM) for its excellent performance [18]. SVM is based on statistical learning theory proposed by Vapnik. It constructs a maximum-margin hyper plane to distinguish the training vectors from different classes in an efficient way [19]. In SVM, maximizing this margin leads to minimization of the probabilistic test error. Furthermore, for non-linearly separable features, a kernel function can be applied. This will map the original problem into a much higher-dimensional space where better classification result may be achieved.

TABLE I.  PERFORMANCE OF THE PROPOSED METHOD IN TERMS OF SENSITIVITY (SE.), SPECIFICITY (SP.) AND ACCURACY (AC.)

|  | Method | Capacity BPB% | SNR (mean±std) | R-MFCC[13] | | | C-R-MFCC | | |
|---|---|---|---|---|---|---|---|---|---|
|  |  |  |  | Se. | Sp. | Ac. | Se. | Sp. | Ac. |
| Steganography methods | Hide4PGP | C=25 | 53.7±7.00 | **100** | 99.5 | 99.8 | 99.7 | **100** | **99.9** |
|  |  | C=12.5 | 65.8±7.00 | 99.9 | 98.5 | 99.2 | 99.6 | **100** | **99.8** |
|  |  | C=6.25 | 72.8±7.00 | 98.7 | 95.1 | 96.9 | **99.5** | **100** | **99.7** |
|  | Steghide | C=3.125 | 74.9±6.45 | 97.9 | 93.1 | 95.5 | **99.5** | **100** | **99.7** |
|  |  | C=1.56 | 77.1±6.05 | 95.6 | 90.2 | 92.9 | **99.3** | **99.8** | **99.5** |
|  |  | C=0.78 | 78.7±5.4 | 92.5 | 87.2 | 89.9 | **99** | **99.7** | **99.3** |
| Watermarking methods | COX | --- | 19.3±3.69 | 95.1 | 98 | 96.5 | **99.8** | **99.1** | **99.4** |
|  | DSSS | --- | 27.7±6.99 | 91.8 | 90.5 | 91.1 | **99.6** | **99.5** | **99.6** |

Non-separable classes can also be distinguished in SVM by using soft margin idea. Soft margin allows hyper plane to split classes as cleanly as possible while maximizing margin between separated samples. In this case, cost of optimization would be a trade-off between large margin and errors due to samples that fall in the margin region. A user defined parameter adjusts effects of these two terms.

The proposed method uses LIBSVM [20] with radial basis function (RBF). Equation (6) shows definition of radial basis function:

$$K(x_i, x_j) = exp(-\gamma ||x_i - x_j||^2), \quad \gamma > 0 \quad (6)$$

## IV. EXPERIMENTS AND RESULTS

### A. Dataset:

The experiments have been carried out on the same database of [21]. It contained a total number of 4169 mono audio signals with sampling frequency of 44100 Hz, resolution of 16 bits, and duration of 10 seconds. The samples were selected from wide range of languages (including English, Persian, Arabic and Turkish) and music genres (including pop, classic, rock, blues, opera, trance, rap, new age, metal and jazz).

The same amount of stego signals were generated by feeding all cover samples into different data hiding algorithms for different capacities. The embedded messages were chosen randomly and each message was used only once. The steganographic methods included Hide4pgp [22] and Steghide [23]. The watermarking methods included direct spread spectrum (DSSS) [24] and DCT-based robust watermarking (COX) [25]. Table I presents details and parameters of the employed database. It is noteworthy that in this paper, the capacity is expressed in term of bit-per-bit (BPB) percent which is defined as ratio of payload size to the cover size in bit. BPB is adopted because not only it quantifies objective of steganography more closely but it is also a universal metric. That is, it can be used across different bit resolutions of covers and it is independent from embedding algorithm.

### B. Discriminative property of features

To provide better insight into potency of the proposed features, a set of typical scatter plots of different settings are presented. Because the graphical plots can be drawn at most in 3 dimensions (which is much less than original space of features), at each experiment feature selection was carried out to find the most discriminative features. Although genetic algorithm has higher search complexity, it can achieve better results. Therefore, feature selection based on algorithm genetic was used for this task. Our GA used a population of 200 individuals with accuracy of the classifier as the fitness function and two point crossover [26] for mating process. The plots are presented in Figure 3. Investigating Figure 3 shows that the proposed features have better discriminative features than their R-MFCC counterparts. Another point that should be mentioned is that, in the feature selection process features corresponding to lower frequencies (near index 1) had lower discriminative capabilities and those corresponding to high frequencies (near index 29) had higher discriminative capabilities.

### C. Performance of the system

Performance of the system was evaluated by dividing database into the train set (70%) and the test set (30%). After training SVM with the first set, it was evaluated with features from the test set. This procedure was repeated for 20 times and the performance criteria were averaged over all repetitions. Performance criteria of the proposed and RMFCC features are compared in the Table I. Furthermore, Table II compares average values of sensitivity, specificity, and the minimum tested embedding capacity ($C_{min}$) of the proposed method and some of previous works.

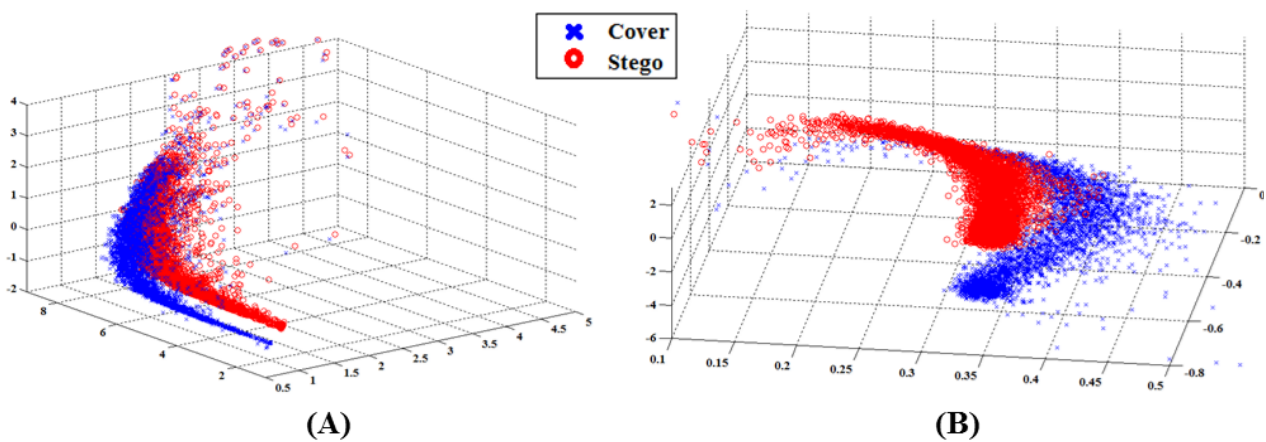

Figure 3. Scatter plots of covers vs. stegos (Steghide@0.78 BPB) (A) R-MFCC features (B) C-R-MFCC features

TABLE II. AVERAGE RESULTS OF STEGANALYSIS IN TERMS OF SENSITIVITY (SE.), SPECIFICITY (SP.)

| Method | Cmin | Se. | Sp. | Ref. |
|---|---|---|---|---|
| AQM | 1.56 | 93.5 | 92.7 | [5] |
| Chaotic | 0.31 | 69.14 | 58.6 | [11] |
| MFCC | - | 66.0 | - | [12] |
| D2-MFCC | 3.156 | AC = 85.9 | | [17] |
| D2-R-MFCC | 1.56 | 94.7 | 97.3 | [13] |
| C-R-MFCC | 0.78 | **99.5** | **99.8** | This paper |

### D. Receiver Operating Characteristic:

Receiver Operating Characteristic (ROC) illustrates how values of true positive and false positive change, as the decision threshold is varied. Figure 4 depicts ROC of different feature sets. Values of area under the curve (AUC) are also reported in Table III.

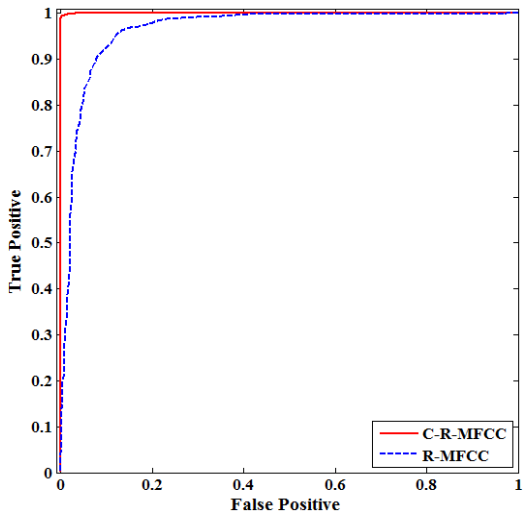

Figure 4. ROC curves of different feature sets for detecting Steghide@0.78 BPB

TABLE III. AREA UNDER THE CURVE (AUC) OF DIFFERENT METHODS (STEGHIDE@0.78 BPB)

| Feature Set | AUC |
|---|---|
| R-MFCC | 0.9527 |
| C-R-MFCC | **0.9999** |

## V. Discussion

Calibration provides estimation about statistical properties of cover signal and can greatly improve performance of steganalysis systems. Referring to figure 3, it is apparent that calibrated features are more separated than their non-calibrated counterparts.

Based on our observation while conducting feature selection we can conclude that the idea of maximum deviation from HAS improves performance of audio steganalysis. According to this idea high frequency regions of signal provide more useful information regarding the presence of hidden messages and thus, these areas should be our main focus for determining the presence/absence of hidden messages.

Finally, consulting results of simulations it can be deduced that the proposed method can detect the presence of hidden messages very reliably, even in low embedding rates. The results from ROC plots also verify that the system provides very good balance between values of false positive and true positive which makes it more practical for steganalysis.

## VI. Conclusion

This paper proposed a new set of calibrated features in order to improve results of audio steganalysis. The calibration was based on re-embedding. In this fashion, reversed Mel frequency cepstrum coefficients (R-MFCC) of both incoming signal and its re-embedded version were calculated. The higher order statistics of differences between these coefficients were used as suitable discriminating features for audio steganalysis. Simulations and analysis were conducted to examine efficacy of the proposed audio steganalysis system. The system achieved very promising and reliable results. Proposed method achieved accuracy of 99.3% (StegHide@0.76% BPB) which was 9.5% higher than previous R-MFCC based method.